\documentclass[runningheads]{llncs}
\usepackage{graphicx}
\usepackage{amsmath}

\begin{document}
\title{State of the Art in Fair ML: From Moral Philosophy and Legislation to Fair Classifiers}

\author{Elias Baumann \and
Josef Lorenz Rumberger}
\authorrunning{E. Baumann, J. L. Rumberger}
\titlerunning{State of the Art in Fair Machine Learning}
\institute{Humboldt University Berlin}
\maketitle             
\begin{abstract}
Machine learning is becoming an ever present part in our lives as many decisions, e.g. to lend a credit, are no longer made by humans but by machine learning algorithms. However those decisions are often unfair and discriminating individuals belonging to protected groups based on race or gender.
With the recent General Data Protection Regulation (GDPR) coming into effect, new awareness has been raised for such issues and with computer scientists having such a large impact on peoples lives it is necessary that actions are taken to discover and prevent discrimination. This work aims to give an introduction into discrimination, legislative foundations to counter it and strategies to detect and prevent machine learning algorithms from showing such behavior.

\end{abstract}
\section{Introduction} 
Big data and modern pattern recognition algorithms together led to a number of remarkable achievements in the last years. Algorithmic techniques surpassed human performance in several tasks, ranging from classic Atari video games \cite{hessel2018multi} or board games like Go \cite{silver2017mastering} to social games like Jeopardy \cite{ferrucci2013watson}. Even more interesting are the successes of algorithms in areas where humans typically dominate, like reading comprehension \cite{yu2018qanet} and image recognition \cite{he2015delving}. Most of the recent advances are accomplished by some kind of specialized artificial neural network algorithm, that show a number of desirable characteristics like universal function approximation capabilities \cite{hornik1989multilayer} and in some configurations even turing completeness \cite{siegelmann1995computational}. However, systems using way simpler algorithms than the above mentioned, are already well capable of beating human experts when it comes to predicting human behavior \cite{dawes1989clinical}. 
\newline
\newline
The rapid development of specialized versions of such algorithms surpassed the construction of theory explaining the reasons for their behavior. However, due to their performance advantages compared to humans, they get used more and more in regular decision making processes. 
Algorithms already govern lots of small decisions in our daily life: the ads that we see online, the composition of our social media feeds and the newspaper articles, books and songs that we get recommended. But it just begun that these systems are used to decide if a child gets a spot at a preferred school, if an prospective student gets the admission to a university, if a entrepreneur gets a loan to start a business or if an individual gets a callback for a job admission \cite{fairmlbook}. Such decisions can have a huge impact on our lives and can determine if we can live up to our full potential or not. Intransparent decisions, potentially flawed with stereotypes and prejudices about certain groups are not acceptable in this process.
Algorithmic decision making promises to be less flawed by the biases incorporated in human decisions and furthermore it could probably find even more nuanced relationships and patters inside the vast inventory of available data than humans ever could. Furthermore those systems get more accessible with the development of frameworks that can achieve high accuracy even without domain knowledge neither about the algorithms nor the task \cite{real2017large}. However, algorithmic decision making is always based on generalizations that arise from past examples and decisions. Thus if past decisions and examples already showed systematic biases like discrimination against certain groups or individuals, an algorithm can not distinguish if that is for a fair reason or just the product of some speculative judgement or worse: targeted mistreatment. Therefore the assumed advantages could vanish and moreover the algorithm can make matters worse: it can obfuscate discriminating decisions by baking them into the more abstract and often undecipherable representation of an algorithm. That is why such behavior is hard to explain and to spot in the first place. Experts are rare and expensive in this field because of its rapid development. Furthermore they can be unaware about the potential discrimination arising from algorithmic decision making and thus these systems can be used for extensive periods without anyone noticing any misbehavior. 
\newline
\newline
In this paper the potential for discrimination induced by algorithmic decision making is assessed and possible methods to cope with such behavior are presented and discussed. First the ethical and moral implications of discrimination and equality are discussed in chapter two. Then the legal situation regarding automated decision making in Germany and the European Union is explained. After that, the main problems and sources of biased decision making of algorithms are laid out in chapter four. Chapter five introduces several formal fairness criteria and points out techniques to develop classification algorithms that satisfy these criteria. The next two chapters present various more advanced methods that look into the inner workings of algorithms and analyze causal relationships in the decision making process. In chapter eight, a discussion is started to reflect the diverse content and gives an outlook on future developments in the field of fair machine learning. The last chapter concludes with some lessons learned.

\section{Discrimination - A Short Introduction}
\begin{quote}
\textbf{Definition:} The use of allegedly clear distinctions of abstract groups to justify unequal treatment that leads to social disadvantages. The imposed disadvantages are not perceived as unjust, rather as a clear consequence of the dissimilarity of the other. \cite{BpB_Disk}
\end{quote}
The authors use the term "abstract group" to denote that people put a sum of other people that do not know each other, but all happen to have a salient feature into a group and judge them based on some kind of assumed collective identity. Examples for such group constructions are "jews", "migrants", "women" and many more that get collective attributes and attitudes assigned. Note that these attributes are not individually assigned based on former experience with individuals. They are rather collectively assigned based on unsubstantiated information or experience with a small sub-sample having the salient characteristic. Disadvantages are always relative to a relevant comparison group. It would be mistaken to refuse discrimination when comparing the treatment of e.g. women in the german labor market with the treatment of women in the labor market of saudi-arabia, because the latter are simply the wrong comparison group \cite{sep-discrimination}. The disadvantages can be of various form, ranging from the exclusion of parts of the social welfare program to the denied entrance to a night club. The education system is an especially sensitive part when it comes to discrimination, since it is often the base for future career possibilities but also a substantial prerequisite to speak out and argue against discrimination. By imposing barriers on educational possibilities for discriminated groups, the discriminating group can impose a sustaining dominance and therefore consolidate its position in society. Discrimination gains traction with rising power imbalance between the groups and does not necessarily reflect any ratio of the number of members between the discriminating and the discriminated group. Discrimination can also have another layer of complexity added when people get assigned to different discriminated groups all at once, e.g. black queer women, which is called intersectional discrimination. This recently really active field of study sheds light on discrimination among discriminated groups and seeks a better understanding of the characteristics of discrimination as an emergent property of the belonging to multiple subgroups.
\newline
\newline
People generally tend to form abstract groups not only for others but also for themselves \cite{allport1954nature}. The group were oneself belongs to is called the ingroup, whereas other groups are denoted as outgroups. Brewer \cite{brewer1999psychology} found that much bias and discrimination is rather motivated by preferential treatment of ingroup members longing to promote and maintain positive relationships among peers than hostility towards outgroups. However, strong ingroup attachment is not necessary for antagonism towards outgroups. Another view is that people compete for rare goods in a society. These goods can be materialistic like housing or idealistic like social recognition and that discrimination is a tool to gain an advantage in the allocation of these goods. The observation that economic crisis typically lead to higher levels of discrimination and xenophobia \cite{bartels2013mass} comply with that theory, since the goods that the members of a society compete for get scarcer during crisis. 
\newline
\newline
Another form of discrimination is Statistical Discrimination, where prejudice free agents behave discriminatory when making decisions under uncertainty and with limited time, effort and information. Often easily observable and thus inexpensive to collect features like age, gender or race are used to infer hard to measure or even unobservable characteristics like ability, motivation or talent. This can result in discrimination of group members who differ from some central tendency of their group distribution of the unobservable characteristic. The group assignment in this case is not based on some visible salient feature and rumors, but rather on the granularity of the available data. Since machine learning (ML) is in context of human characteristics exactly used to infer hard to measure or even unobservable features, it could be prone to statistical discrimination. 
\newline
\newline
\label{obsApp}
Since ML is nowadays used to make important decisions about peoples fate, it is important to detect discriminating behavior of such algorithms. A problem imposed especially by Artificial Neural Network based algorithms is, that the trained algorithm itself is a non-interpretable highly complex non-linear function, therefore often referred to as a black box. Hardt et al. \cite{hardt2016equality} propose the observational approach that circumvents the black box problem: In an ceteris paribus experiment, an algorithm under analysis is fed with some data from a possibly discriminated group, e.g. women, to predict some label. Then the group membership is switched to the reference group, here male, and the difference in outcome is analyzed. In that way, one can infer possible causal relationships between the belonging to a possibly discriminated group and its effect on the prediction made by an algorithm.   
\subsection{Equality of Opportunity (EoO)}
Researchers from such diverse disciplines as economics, philosophy, computer science, sociology and many more work on topics related to EoO. Therefore the context in which EoO is analyzed often encompasses the distribution and allocation of goods, services, educational prospects and job opportunities. 
\subsubsection{Formal Equality of Opportunity:} \label{formalEoO}
Two notions of EoO are reported, the weaker is called Formal Equality of Opportunity and is satisfied when people, based on all relevant aspects for a certain task or study program have the same chance to be taken and thus are treated equal \cite{sep-equal-opportunity}. According to the weak notion, two students with the same grades in the same undergraduate program (e.g. CS) should be treated equal when it comes to the graduate school admission for a CS program. Another prospective student exists with an equal skill level in the relevant field but who has a major in another subject, say economics, and applies to the CS graduate program. In a perfect formal EoO world, the student would have the same chance as the two other applicants. However in the real world, proxy variables such as the grades in specific courses are used to build a prediction about the ability. Thus an applicant who has no certificate about the specific courses might get rejected. Problems like this can be circumvented by higher efforts from the admission office, like standardized tests or interviews to check for motivation and ability, but they all come at a higher cost for both the admission office and the applicant. Statistical discrimination as explained before violates formal EoO too, since irrelevant aspects like specific certificates or cultural heritage are used as predictors for relevant but expensive to obtain or unobservable aspects. Both examples show, that one weakness of formal EoO is the definition of relevant aspects. Especially in an discrimination context, there have been times in almost all countries of the western world, were people were sure that certain minorities of the population are not capable of the same achievements as the rest, due to some inevitable different trait verified by some pseudo-scientific method. One could use this verifiable short coming as an entry requirement to impose a barrier for that particular minority. Another blind spot of formal EoO is that discriminated individuals can have worse observable relevant aspects due to long lasting structural inequality, even though these individuals have the same level of motivation, eager and curiosity as their non-discriminated peers.
\subsubsection{Substantive Equality of Opportunity:} \label{SubstEoO}
Therefore the broader notion of EoO, called Substantive Equality of Opportunity is satisfied when individuals with same talents and ambition can achieve the same outcome in life \cite{sep-equal-opportunity}. This encompasses that for individuals belonging to socially disadvantaged groups, the game of taking part in social participation is probably rigged from the moment when their socially disadvantaged parents gave them birth. This is especially interesting in context of high social immobility in germany compared to other northern european countries \cite{madoc35059}. Substantive EoO supporters advocate for the state to step in and fill the gap by providing healthcare and educational possibilities especially for children of socially disadvantaged families \cite{krugmanEoO}.
\newline
\newline
Affirmative action is another technique with the goal to reach higher substantive EoO and is mostly pursued in the US and UK but probably will have an impact on german assessment practices too. The concept is to uplift socially disadvantaged groups to a level playing field to break the cycle of sustained discrimination. This must be done as early as possible since the first certificates from the education system already define a track for future educational prospects. It can be accomplished by providing resources through redistribution of taxes or transfers from rich neighborhoods to poor ones and some other redistributional techniques only the state can employ. Other forms are the introduction of quotas for specific university programs or job positions like the german female quota, which obliges big companys to have 30\% women in their supervisory board. The overall goal of substantive EoO and therefore affirmative action is that every individual in a society has the possibility to reach to their maximum potential. However affirmative action policies are under criticism, because they sometimes contradict formal EoO and directly discriminate someone based on their race. A good example is the case of Michigans university admission practice that had a quota on social disadvantaged groups based on racial background. The quota led to the admission of students belonging to a disadvantaged group which had worse grade point averages (GPAs) in their high school diplomas than other candidates that got rejected. In 2003 it was challenged by a prospective student that got rejected, despite her superior GPA. The US Supreme Court ruled in 2003, that the universities practice to favor disadvantaged groups over the rest is legal. In 2006 a poll among Michigan voters decided to abolish the practice which was followed by several legal proceedings which are in ongoing trials about the policy \cite{nytAffirAct}.
\subsubsection{John Rawls' Theory of Justice as Fairness:} 
In the liberal political conception, legitimacy of a political system is only the minimum in terms of moral standards. This intuitively makes sense in context of Max Webers idea that reign can be sufficiently legitimated  by success \cite{weber2002wirtschaft}. For example an immoral system, that brings economic advantages to broad parts of the population under control can be legitimated by its people if they just roughly behave economically rational. Moral philosopher John Rawls describes justice in context of political systems as an arrangement of institutions that are morally best and therefore sets justice as the highest achievable state in terms of moral standards. Rawls then defines justice as fairness among the constituents of a society and splits it into the following two principles \cite{sep-rawls}:
\begin{enumerate}
    \item First Principle (Liberty): Every individual in a society has some basic reciprocal rights like freedoms of expression or conscience, political participation and property rights
    \item Second Principle: Social and economic inequalities have to satisfy two conditions:
    \begin{enumerate}
        \item Substantial Equality of Opportunity is satisfied in the participation mechanism gatekeeping the advantages of these inequalities.
        \item Difference Principle: Inequalities are only accepted if they work to the advantages of the most disadvantaged groups in society. 
    \end{enumerate}
\end{enumerate}
An example for the difference principle is the high salary of physicians compared to say workers. The income inequality makes sure that a reasonable number of people wish to become physicians and a higher number of physicians is beneficial for the whole society. The difference principle is also satisfied by most affirmative action policies, since the introduced inequalities typically benefit the most disadvantaged groups. 
\newline
\newline
Rawls came along with these ideas by applying a simply yet elegant thought experiment called the original position. In this framework, it is assumed that several individuals start at an hypothetical position where they are not part of a society, but will be part of it in the future. The key element is that they do not know in the first place into which part of the society they will be born. So they do not have information about their future gender, geographic origin, cultural background or anything else that could possibly lead to inequalities. In this artificial position they have to negotiate the basic structures, duties and prospects of a society. Since anybody has the chance to be born into any part of the society, Rawls assumes that people would negotiate fair social institutions for anyone in the society. 
\section{Legal Situation - Germany}
The german legislative introduced in 2006 the General Act on Equal Treatment (GAET) where the following attributes are defined as protected:
\begin{itemize}
    \item ethic or national origin and language
    \item gender or pregnancy
    \item religion or belief
    \item disability, chronic or mental illness
    \item age (both, old and young)
    \item sexual orientation
\end{itemize}
People having one of these attributes are protected from discrimination by the legislation. \S 2 GAET explains the scope: legal entities have to comply with the legislation in relation to conditions for the access of employment and promotion, employment conditions like pay,reasons for dismissal, vocational training and memberships and benefits of workers or employers organizations. Furthermore they have to be treated equal when it comes to social security or other social advantages, education and access to goods and services that are available to the public, including housing. Nevertheless legislation referring to contractual freedom exists too and makes it possible that people can refuse to make contracts with individuals without specifying a reason. In \S 5 GAET the legislation refers to positive action (i.e. affirmative action) and states that unequal treatment is only allowed, when it is appropriate to prevent or compensate disadvantages against protected groups. \S 8-10 GAET introduce additional limitations for the equal treatment specified in \S 1-2 GAET, however these are  not related to the topic of this paper.
\subsection{General Data Protection Regulation (GDPR)} 
The European Union (EU) adopted the GDPR \cite{gdpr} in 2016 and it became enforceable in May 2018. The objective of the regulation is the protection of individuals in relation to the processing and movement of their personal data. Therefore it introduces a fundamental right to the protection of personal data. In the following some key sections concerning the topic of this paper are presented:
\newline
\newline
Article 4 GDPR gives several broad definitions about the terms used in the document, e.g. personal data is defined as any information related to identifiable individuals, which could be identified via name, ID, location or other online identifier. Most definitions are broad in the sense that they rather span too much than to leave a definition gap and therefore allow data collectors to circumvent the regulations. 
\newline
\newline
Article 9 GDPR forbids the processing of personal data that reveals the membership to one of the protected groups. The protected groups are essentially the ones specified by the GAET and some more like membership in a trade union. Furthermore article 9 GDPR makes it illegal to process without explicit consent genetic or biometric data for the purpose of identifying an individual and data concerning health or a persons sex life.
\newline
\newline
Articles 12-20  GDPR gives individuals the right to access their personal data if collected by some legal entity. The data collector has to provide a concise and accessible explanation of the data collected, the purpose of the processing and the disclosure of data to third parties. The logic governing profiling practices must be explained in an understandable way. Furthermore the planned length of storage of the data has to be noted and the individual whose data has been stored and/or processed (i.e. data subject) has the right for rectification or erasure of its data on the storage of the data collector. 
\newline
\newline
Articles 21-22 GDPR gives an individual the right to object against automated data processing and decision making. After objection, the data collector is no longer allowed to process or base decisions concerning the individual on automated processing.
\newline
\newline
Article 35 GDPR obliges data collectors to assess the risks to the rights and freedoms of individuals caused by their data collection and processing practices prior to their deployment. The process is called Data Protection Impact Assessment and has several official requirements and data collectors shall seek advice from official data protection officers when in doubt.
\newline
\newline
Additional to the regulations in the GDPR and GAET, article 21 of the Charter of Fundamental Rights (CFR) of the EU adds the social origin and property of an individual to the protected groups. All regulations are probably sufficient for formal EoO and make certain steps towards substantial EoO via the legalization of affirmative action and the protection based on social origin. However the CFR is just enforced in rare cases where a national legislation breaches it or when an protected group is discriminated by the executive of a member state of the EU.
\section{Discriminatory behavior of algorithm}
For better accessibility, the next sections will analyze a reduced classification problem where features $X$ are used to predict label $Y$. The predictions of the classifier are denoted as $\hat{Y}$. The input features $X$ contain explicit or implicit (i.e. latent) one or more protected characteristics $A$. A protected characteristic can be for example any of the characteristics mentioned in \S 1 GAET. A real world example for this classification setting could be a company, which seeks to process personal data $X$ from their employees. Besides a variety of other information, the data set contains information about the gender $A$ of each individual. The goal of the company is to replace decisions about job promotions made by managers with ones made by an automated classifier to improve fairness. Job promotions were either granted ($Y=1$), or denied ($Y=0$). Promotion decisions $Y$ are used together with data $X$ to optimize a classifier, so that next years promotion decisions $\hat{Y}$ can be determined using the algorithm. Since the company wants to have a fair algorithm and knows about the potential for discrimination that could arise from including protected characteristics like an individuals gender, the data scientist decides to remove feature $A$ from the data set $X$. Unfortunately this practice of \textbf{fairness through unawareness} usually does not make the algorithm more fair and can even be harmful \cite{fairmlbook}. Rich data sets like $X$ in our example, typically contain a number of features which are slightly correlated with protected variable $A$. Therefore these features can together be used to predict the protected variable. If $A$ is omitted, these other features correlated with $A$ will still carry the information about a protected group membership and therefore the model does not become fairer.
To highlight the case even further, Sugimoto et al. \cite{sugimoto2016everything} make the case that everything might reveal everything, in the sense that given a sufficiently large feature space, one can probably infer all unknown information about an individual. Some examples for this are presented in the following:
\begin{itemize}
    \item Danford and Reece \cite{reece2017instagram} apply machine learning methods on instagram photos to predict the probability of an individual to suffer from depression. The resulting model beats average human practitioners in unassisted diagnosis accuracy. Suffering from mental illness is a protected variable under german legislation.
    \newline
    \item Kosinski and Wang \cite{wang2017deep} studied a pre-trained face detection algorithm (VGG Face), which is widely used by companies and governments. They applied it on facial images from an online dating platform and showed that the embedding space of the pre-trained algorithm encodes features that allow for predicting the sexual orientation of individuals with higher accuracy than humans. By applying a simple logistic regression model on the VGG Face embedding space, they were able to determine if an individual is gay or not based on a single photo in 81\% of the cases for males and 71\% for females respectively. Human accuracy for the same task is just 61\% for men and 54\% for women on average. Given that homosexuality is illegal in several countries and an important reason for unequal treatment in many others, the authors state that their findings expose a threat to the privacy and safety of homosexual men and women.
    \newline
    \item Fiscella and Fremont \cite{fiscella2006use} review methods to estimate race and ethnic background based on surnames and geolocations of individuals.
    \newline
    \item Narayanan and Shmatikov \cite{narayanan2008robust} present several statistical methods to de-anonymize large sparse datasets. They apply these methods to de-anonymize individuals from the Netflix Prize dataset based on information about users from the comments and ratings in the publicly available Internet Movie Database (IMDB). The former dataset was made public by netflix in order to organize a data science competition. After successful de-anonymization a lawsuit against netflix was filed and the data science competition was discontinued due to privacy concerns. The authors found that besides real identities, they could uncover apparent political preferences from the data.
    \newline
    \item De Montjoye et al. \cite{de2013unique} show that on average four spatio-temporal points (geolocations with time tags) are sufficient to identify 95\% of the individuals in their dataset containing coarse mobility information about one and a half million individuals during fifteen months. Since many applications on smartphones or other devices require users to send current geolocations, one can imagine the ease of de-anonymization. 
\end{itemize}

\subsection{Where does discriminatory behavior of algorithms come from?}
Discriminatory behavior of algorithms can arise from several different factors. The process from data generation and collection via pre-processing to inference, post-processing and interpretation often involves the contributions of several individuals often working in different departments or even different companys. Therefore these factors are not obvious and easy to spot for any of the individuals in the chain. In the following section some common problems that lead to algorithmic bias against protected groups are explained. Since there could be numerous other reasons for bias, this should not be regarded as a exhaustive list.
\subsubsection{Selection and Confirmation Bias:}
 Lum and Isaac \cite{lum2016predict} analyzed algorithmic systems for the allocation of patrolman, so called predictive policing systems, used by the police department of Oakland, CA. Police officers used to allocate patrolman into neighborhoods with high numbers of immigrants and low social prosperity due to biased decision making. The predictive policing system learned that pattern and therefore proposed a similar allocation practice. Since the patrolman were predominantly patrolling in quarters where people with immigration or low social background live, they did more police checks on these people than on others. This is called selection bias, since the police already focuses on a certain subset of the population. The data that is generated by this practice reflects the pattern, that more crime happens in these quarters than elsewhere and therefore supports the allocation decision made by the predictive policing system. This support for the already biased selection is called confirmation bias and can result in self-sustaining cycles of discrimination. The authors of the study found that black people are roughly targeted twice as much as white people for drug controls. Even though official statistics show that both groups show roughly equal rates of drug abuse.
\subsubsection{Limited Features:} 
\begin{figure}
    \centering
    \includegraphics[width = \linewidth]{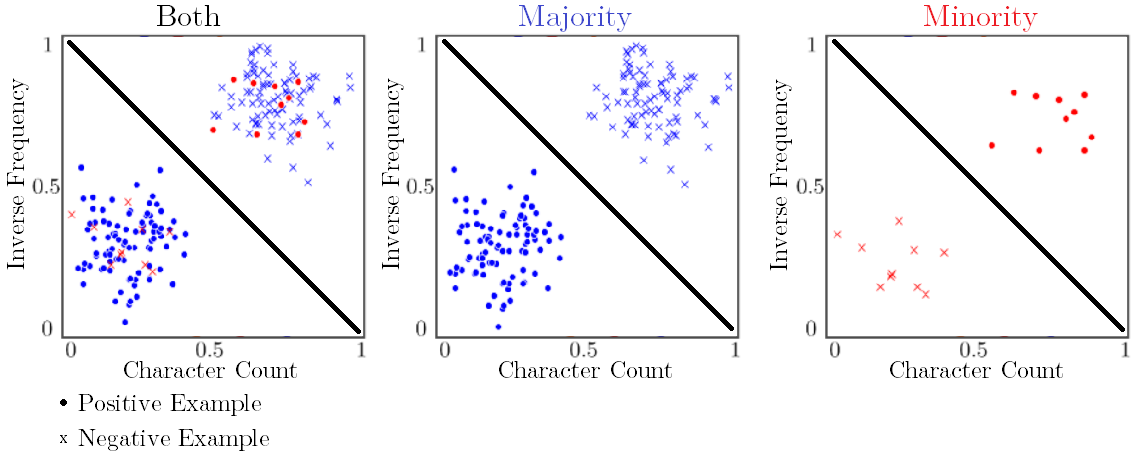} 
    \caption{Classifier trained to separate real names (positive examples) from pseudonyms}
    \label{fig:sampleDisp}
\end{figure}

Given a dataset containing personal data about a population with a big proportion belonging to a homogeneous cultural majority and a minority coming from diverse cultural backgrounds. Features that have high explanatory power for the majority can have significantly lower or even reversed explanatory power for the minority. Figure \ref{fig:sampleDisp} illustrates this for the example of name checks, to separate real names (positive examples) from pseudonyms (negative examples). On the x-axis is the number of characters in a given name on a normalized scale and on the y-axis is the inverse frequency of a given name in the dataset. A classifier was trained on the data and its decision boundary is shown as a black line separating the real names from the pseudonyms. In our example the majority consists of western surnames like "Cathy Amber", "Mike" or "Joanna" and the minority of southern indian surnames like "Pilavullakandi Thekkaparambil" or "Villupuram Chinnaiahpillai". The classifier showed $90\%$ accuracy overall. If one takes a closer look at the predictions, it becomes clear that the algorithm has $100$ \% accuracy on the majority, but $0$\% accuracy on the minority and therefore these features are infeasible to distinguish between real names and pseudonyms for both groups at once. Problems like this were reported during the 2011's "Nymwars" where Facebook and Google+ tried to enforce a real-name policy on their platforms by banning accounts that used pseudonyms. 
\subsubsection{Sample Size Disparity:} 
Buolamwini and Gebru \cite{buolamwini2018gender} analyzed several commercial APIs from Google, Amazon and Microsoft that provide a gender recognition service by analyzing images via computer vision algorithms. The authors found that the software worked remarkably good for images of white individuals, with accuracy rates up to $99$\% for white males and $93$\% for white females. When fed with images of black individuals the accuracy dropped down to $86$\% for black males and $70$\% for black females. The authors investigated that the used training datasets consisted of highly unbalanced samples of the population with up to $86.2$\% individuals of lighter skin color. Therefore the algorithm preferably learned patterns that helped to distinguish white males from females but are not good applicable on black individuals. Sample size disparity can make the problem of limited features more severe.


\section{Discrimination detection and prevention with the observational approach}
The observational approach, already mentioned in chapter \ref{obsApp}, uses easily observable characteristics like the features $X$ including the protected variables $A$ and the target labels $Y$ and their predictions $\hat{Y}$. By analyzing conditional probabilities of $Y$ or $\hat{Y}$ given $X$ and $A$, one can spot discriminating behaviors without taking a look into the inner workings of the classifier employed on the data \cite{hardt2016equality}. In practice the observational approach can be used to test classifiers of any form by applying a simple ceteris-paribus experiment: the classifier gets fed with two datasets: an original dataset and one where protected features or proxy variables for these variables are swapped from a protected group to an unprotected one and vice versa, holding the rest of the features fixed. After obtaining the classifications $\hat{Y}$, one can analyze the effects of belonging to a certain protected group on the predictions $\hat{Y}$. Bertrand and Mullainathans provide a good example \cite{bertrand2004emily}: they sent out identical fictitious job application documents ($X$) to a number of companies and use different names, suggesting the gender and the ethnic background of a candidate ($A$). By doing so, they found that fictitious candidates with white sounding names receive on average 50\% more callbacks for job interviews ($\hat{Y}$) compared to identical candidates with black sounding names. The strength of the observational approach is its applicability on any classification process, automated or not, if the required observations are available. The observational approach then analyzes the accuracy, sensitivity and specificity of the predictions made by the algorithm and uses them to infer fair classification post-processing routines.
\newline
\newline
Since fairness is a debatable topic, there exist a number of papers that propose different kinds of fairness criteria. The group fairness criterion, also called demographic parity, is satisfied when positive classifications $\hat{Y}$ are statistically independent from protected variables $A$. In other words: the percentage of individuals in a protected group which are classified as positive have to be the same as the overall percentage of all individuals. On the other hand there exist two different individual fairness criteria which are satisfied when similar individuals get a similar classification. The weaker form called equal opportunity just focuses on positive classifications, whereas the stronger form called equalized odds focuses on all classifications and misclassifications. Each has to be equal when comparing the protected group to the rest of the population. The US Equal Employment Opportunity Commission (EEOC) uses a relaxed form of the equal opportunity fairness criterion to spot unequal treatment. They apply the so called 80\% rule, which allows a 20\% difference in treatment of the protected group compared to the rest.

\subsection{Literature regarding the observational approach}
One branch of papers seek to encode the original data into a fair representation, such that the protected group can no longer be inferred while all other information is retained as good as possible. Zemel et al. 2013 \cite{zemel2013learning} propose an approach that satisfies both, demographic parity and individual fairness after the application of such an encoding. They write and train an algorithm to both obfuscate the protected variables as well as preserving all relevant information to retain high accuracy when using this encoded dataset for classification. Despite demonstrating decent results, they report that their approach often results in negative effects for different groups as more false positives from one group might be used in order to achieve parity. Feldman et al. \cite{feldman2015certifying} propose the balanced error rate as an accuracy measure that can be used to construct an unbiased data representation. However, classifiers trained with this unbiased data representation can exhibit the same behavior of adding false positives from the disadvantaged group. Therefore the authors propose several methods to cope with that problem by only partially changing the data set to construct an unbiased representation. Furthermore they develop a likelihood-ratio based test, that can be used to detect unequal treatment that exceeds a predefined maximum. This test can then be directly used to test if a classifier complies with the aforementioned 80\% rule required by the EEOC. Edwards and Storkey \cite{edwards2015censoring} use adversarial learning to build a fair representation of a cross-sectional dataset, but also to obfuscate private information like names or number plates from images. Adversarial learning is a two step approach: an encoder is used to encode the original dataset into a fair representation and an adversary classifier uses this data to predict the protected variables. Both learning algorithms are trained, until the classifier cannot distinguish any more if data came from a protected group or not. The authors extend that model by introducing a decoder that tries to reconstruct the original dataset from its fair representation to make sure that the dataset is changed as little as possible while achieving the goal of constructing a fair representation. Madras et al. \cite{madras2018learning} extend that model by introducing an adversarial that can incorporate each of the aforementioned fairness criteria. Additional to that, they derive theoretical worst-case guarantees of the constructed fair representations, which makes their method especially interesting for practitioners facing compliance risks.
\newline
\newline
Another more recent branch of literature advocates a completely different technique, that does not construct a fair representation of the dataset. Instead the original dataset is used to construct a classifier that predicts probabilities. Since the objective of the classification problem is a binary outcome, the probabilities have to be rounded up or down at some threshold.
Zafar et al. \cite{zafar2015fairness} show, that by using different thresholds, one can achieve different degrees of fairness among the population. Again a classifier that complies with the EEOC's 80\% rule can be constructed. Hardt et al. \cite{hardt2016equality} extend this approach and introduce a method to construct thresholds that shift the cost of misclassifications from the disadvantaged group to the decision maker. Furthermore they present a detailed analysis of the limitations of the observational approach.
In the following sections, the fairness criteria are thoroughly examined and thresholding techniques are explained.

\subsection{Demographic parity} 
Demographic parity as mentioned before is satisfied when the positive predictions of a classifier $\hat{Y}$ are statistically independent from the membership of a protected group $A = a$. 
\begin{equation}
    P(\hat{Y} = 1 | A = a) = P(\hat{Y} = 1 | A = b)
\end{equation}
A positive classification here does not necessarily have to be beneficial for the individual i.e. could also be a rejection inside an application process. The implementation of the fairness criterion is convenient, because it is not dependant on the true values $Y$. The true values $Y$ are in many cases really hard to define, to observe and to measure. When it comes to job applications for example, what is it that the employer is looking for in a candidate and how is it measured? Often these true values $Y$ are just noisy approximations, sometimes biased with personal beliefs of individuals in the data generating process \cite{fairmlbook}. Demographic parity is not sufficient to satisfy formal EoO as specified in chapter \ref{formalEoO}, because it would not provide equal opportunities based on an individuals observable characteristics. In fact it would give an advantage to individuals from the protected group when it comes to positive classifications if the protected group is on average worse in the observed features than the rest. Because of that, demographic parity can be used as a fairness criterion to implement an affirmative action policy. That could leverage the opportunities of socially disadvantaged groups and therefore might be a step towards substantial EoO (chapter \ref{SubstEoO}).

\subsection{Equal opportunity and equal odds} 
Demographic parity has a number of issues to where it cannot be considered a fair adjustment to any decision. Furthermore, as demographic parity requires the target and protected variables to be un-correlated, any instance where this is the case cannot be predicted perfectly anymore \cite{hardt2016equality}. An example for the necessity of this measure would be a job hiring process for software engineers, where job advertising should be seen by the right individuals. Here it is likely that more men are targeted than women, strictly because there exist more men than women in the profession of software engineer. 
Considering these issues, Hardt et al. 2016 \cite{hardt2016equality} propose two new fairness measures, equalized odds and equal opportunity. 

\subsubsection{Equalized odds:}
 This fairness criterion refers to the prediction $\hat{Y}$ being independent of $A$ conditional on $Y$. For a binary classifier this means the constraint requires the classifier to have equal true and false positive rates among all protected groups \cite{hardt2016equality}. Equation~\ref{eq:sepa} demonstrates equalized odds as two conditional probability statements (1. True postives, 2. false positives).
\begin{equation}
\begin{split}
P(\hat{Y} = 1|Y=1,A=a) = P(\hat{Y} = 1|Y=1,A=b) \\
P(\hat{Y} = 1|Y=0,A=a) = P(\hat{Y} = 1|Y=0,A=b)
\end{split}
\label{eq:sepa}
\end{equation}
Using the previous example of software engineer hiring process, the percentage of women and men receiving an ad who are not software engineers has to be the same and the percentage of women and men receiving an ad who are software engineers have to be the same.

\subsubsection{Equal opportunity:}
The other concept proposed by Hardt et al. 2016 \cite{hardt2016equality}, called equal opportunity, is a less strict version of equal odds, where only the beneficial outcome has to be equal for all protected groups. Therefore the same conditional probability holds, except the target being the constant beneficial outcome. This can be compared to equal true positive rates for all groups. Again, referring to Equation~\ref{eq:sepa}, equal opportunity only requires the first row to hold, not the second.
Relating this to a credit lending example, the percentage of individuals who should receive a credit as they are able to pay it back, should be the same for all protected groups.
\newline

\begin{figure}
    \centering
    \includegraphics[width = \linewidth]{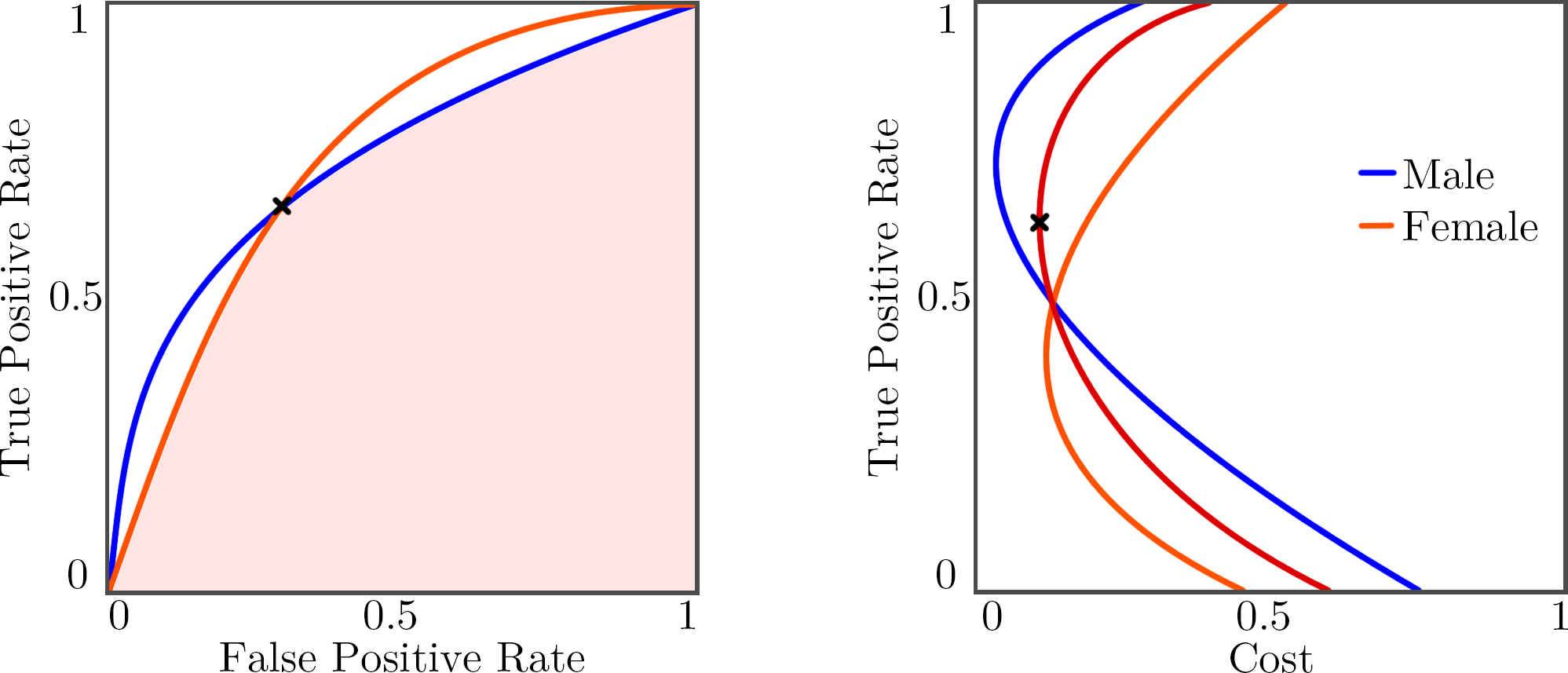} 
    \caption{ROC Curve and True Postive Rate to Cost curve with equal odds / equal opportunity selection. (Curves were created for this example)}
    \label{fig:rocthresholding}
\end{figure}

For both measures, this holds for binary and continuous classifiers, as for continuous classifiers the conditions need to hold for any threshold to satisfy equal odds or equal opportunity. Implementing equal odds or equal opportunity can be done by correctly thresholding a classifier such that the above conditions hold. This can be explained in a simple way using receiver operating characteristic (ROC) curves. ROC curves can show the false positive rate in relation to the true positive rate depending on the set threshold. Figure~\ref{fig:rocthresholding} shows optimal thresholds using curves of two different protected classes (e.g. male, female). If equalized odds has to be achieved, any point in the graph can be taken where all curves intersect. As it is likely that there exists no such point, the threshold has to be set optimally for the lowest curve and all other curves have to take on a higher false positive rate. Considering Figure~\ref{fig:rocthresholding}, if the curves did not intersect, any point in the light red space can be considered. This means that for all groups where the chosen threshold is not on the curve, there is a manual increase in false positives which again produces similar issues as demographic parity.

If only equalized opportunity is considered, the curve is instead drawn as true positive rate to cost and it is attempted to minimize the average cost while maximizing the average true positive rate. Cost here can either be the actual financial cost for a business or a more general social cost for the specific groups.

However as critizized in other works \cite{zafar2017fairness}, this approach requires information on all protected variables of every individual. Hardt et al. \cite{hardt2016equality} themselves also report that this does not solve the issue of a dataset having biased labels.
Zafar et al. \cite{zafar2017fairness} demonstrate a similar approach they call disparate mistreatment. A classifier suffers from disparate mistreatment if either false positive or false negative rates are not equal among protected groups. They weaken their constraint saying that in specific applications such as incarceration one might only require false negatives to be equal. Here their argument would be that it is worse to falsely incarcerate somebody than to falsely let sombeody go.
The constraint is very similar to \cite{hardt2016equality}, however they propose a convex minimization problem which classifiers should optimize in order to train the algorithm not to include disparate mistreatement. This differentiates them from the work of Hardt et. al as the classifier can then be used for new examples whereas Hardts approach post-processes predictions.
For both approaches another issue is the availabilty of ground truth. Both methods require that the data includes ground truth, e.g. that a person has actually commited a crime or is actually a software engineer or not. In cases like criminal classification, the information whether a person actually committed a crime might never be available.
Both individual fairness criteria can satisfy formal EoO, if all sources of bias in the data collection process are ruled out. Substantial EoO could in theory be achieved, if a classifier is fed with a data set that includes strong proxies for ambition and talent, while omitting any data that is generated by the biased processes a socially disadvantaged individual has to cope with.

\subsection{Calibration} 
\begin{figure}
    \centering
    \includegraphics[width = .6\textwidth]{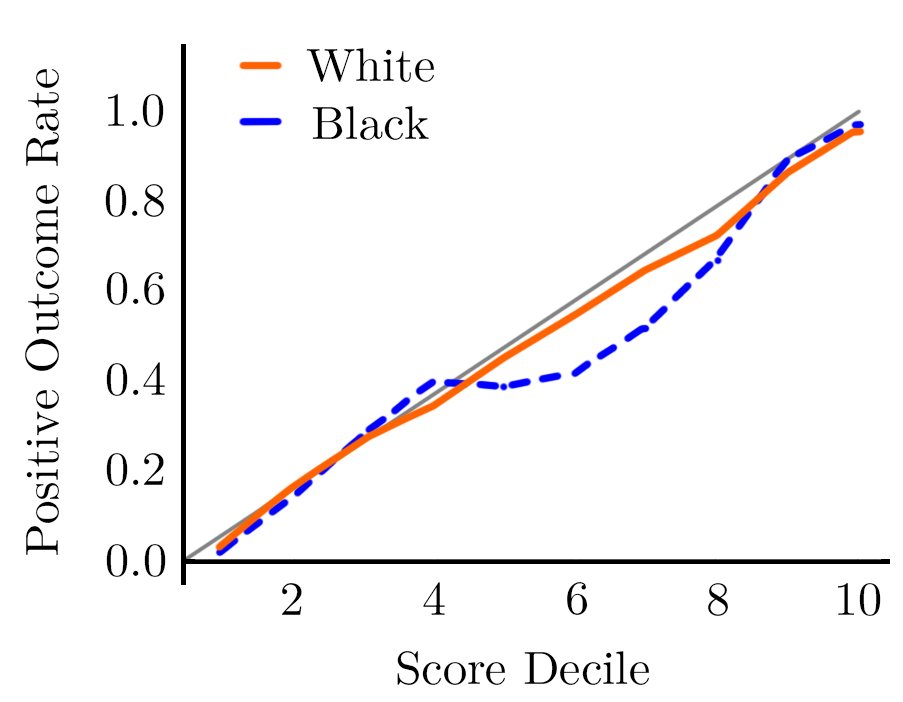}
    \caption{Uncalibrated classifier. Logistic regression on income predictions using UCI Adult Census Dataset \cite{Dua:2017}}
    \label{fig:calibration}
\end{figure}

A third observational approach, often simply referred to as calibration, is a fairness criteria which in many cases already holds as algorithms optimize for it to hold if they are able to observe the sensitive variables \cite{fairmlbook}. In their book, Barocas et al. \cite{fairmlbook}, they call this criterion sufficiency, as the classifier $ \hat{Y}$ is sufficient for predicting the target $Y$ and it incorporates the protected variable $A$ (see equation~\ref{eq:suff}).
\begin{equation}
    P(Y=1|\hat{Y}=\hat{y}, A =a) = P(Y=1|\hat{Y}=\hat{y}, A =b)
    \label{eq:suff}
\end{equation}
A correctly calibrated classifier supports this notion as for every class of A holds:
\begin{equation}
    P(Y=1|\hat{Y}=\hat{y}) = \hat{y}
\end{equation}
To understand calibration, consider figure~\ref{fig:calibration}. We see the relation of probability of being in group $Y=1$ and the percentage of the group being in $Y=1$ for that specific probability.
An optimally calibrated classifier here would always attempt to have the percentage of individuals with a specific score to be equivalent to that probability score. Therefore the optimally calibrated classifier would have all curves on the diagonal. Calibration methods are forms of post-processing, where the ground truth is not required \cite{fairmlbook}. To calibrate an algorithm, a model such as logistic regression and the prediction and outcome values for every group is considered. This addtional model attempts to move the curve closer to the diagonal. An example for this technique is Platt-scaling which was originally designed for support vector machines but can be applied to any classification problem \cite{platt1999probabilistic}.
This type of fairness is also picked up by Chouldechova \cite{chouldechova2017fair}. They additionally argue that predictive parity is mutually exclusive with calibration by group and therefore for every specific task, the impact of applied fairness criteria has to be understood. However not only these two fairness criteria are  mutually exclusive which leads us to the problems and limitations of the observational approach.

\subsection{Limitations of the observational approach}
Even though all three criteria for fairness and non-discrimination cover a different type of fairness, no two of the criteria can be applied at the same time. For the first criteria, demographic parity, fairness is achieved by disregarding the protected attribute but both other criteria explicitly make use of the protected variable and allow for dependencies between protected variables and classifier. However, equalized odds and equal opportunity are also in conflict with calibration methods as proven by \cite{chouldechova2017fair}.
Because of the exclusivity, practitioners have to choose which fairness criteria to apply. This in turn can be difficult as it may lead to additional discrimination \cite{kusner2017counterfactual}. 
Specifically equal odds and demographic parity may lead to an increase in false positives for some protected groups, which in turn for example could lead to an increase in hired software engineers who are no software engineers. This hiring of incompetent individuals will in turn be noticed by individuals of other groups and can then be taken as a simple discriminating argument against the entire group being viable for the software engineer position. This effect can to some extent be observed in the womens quota in germany, where radical opponents of the quota use any false positive as a reason for outrage (see e.g. \cite{womenquota}).
Another issue is the disregard of long-term impact of decisions, because fairness at decision time can have negative impact in the long run. Take for example credit lending, where we could apply equal false and true positive rates for all groups. However a false positive here means that the person will default in the future leaving them with financial ruin.
Finally, considering that individuals can have potentially discriminating intentions, by knowing the criteria, they can influence the gathered observations such that the intended discrimination is reflected perfectly in the dataset.

\subsection{An impossible example}
\begin{figure}
    \centering
    \includegraphics[width=\textwidth]{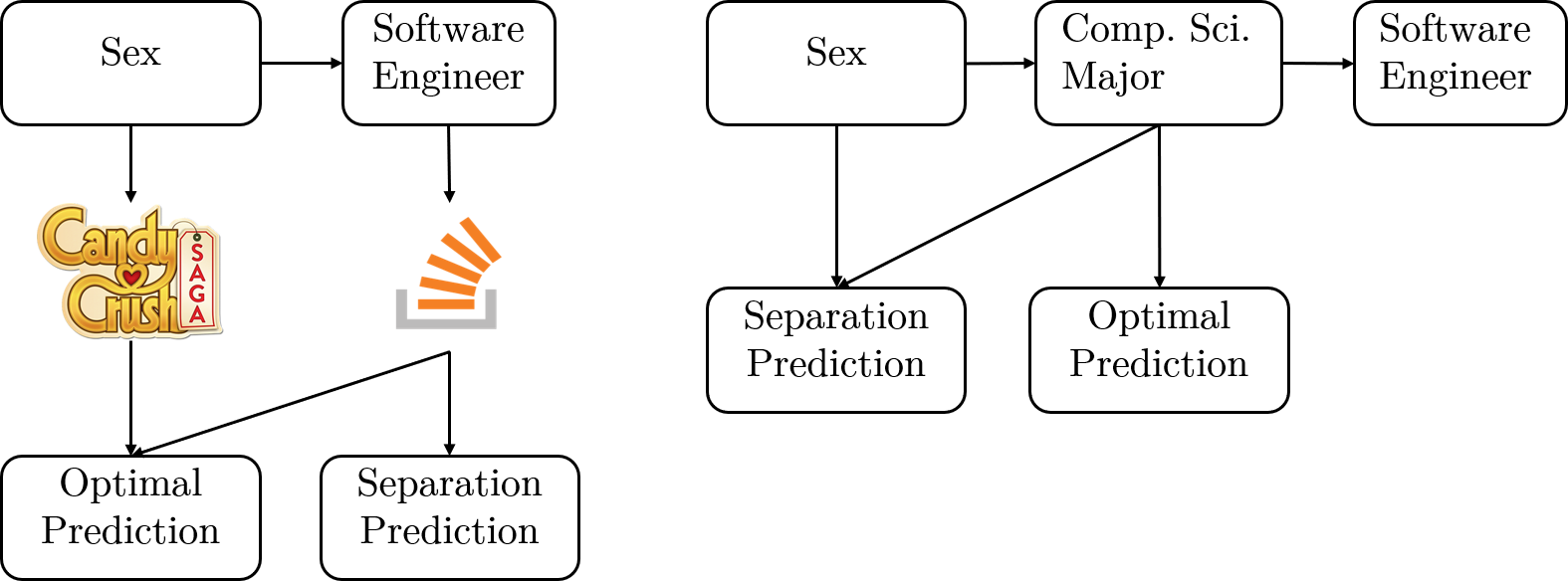}
    \caption{Indistiguishable examples for observational criteria}
    \label{fig:limobs}
\end{figure}

To introduce the next chapter on causal reasoning, we will construct an example based on \cite{kilbertus2017avoiding,hardt2016equality} with two rather different scenarios from a fairness standpoint which yet behave equal when considering observational criteria. The scenarios in this example are again related to the example of showing software engineering job advertisments to men and women. Figure~\ref{fig:limobs} shows causality graphs for those scenarios, where arrows indicate correlation or that $A$ in parts produces $B$ and boxes represent variables and classifiers.

\begin{enumerate}
    \item The first scenario in Figure~\ref{fig:limobs} uses biological sex as a protected variable, the binary indicator wether a person plays the videogame Candy Crush, the binary indicator wether somebody uses the popular programming support site Stack Overflow and the target wether an individual is a software engineer (binary as well). The optimal prediction in this scenario would take both Candy Crush and Stack Overflow use into account because Candy Crush is a weak proxy for biological sex and sex and software engineer are correlated. Stack Overflow is directly generated from software engineer and therefore a good indicator by itself. The predictor using equalized odds would refrain from using Candy Crush and stick to Stack Overflow as then the conditional independence of the predictor as seen in Equation~\ref{eq:sepa} holds.
    \item The second scenario in Figure~\ref{fig:limobs} again uses biological sex as the protected variable and binary variable wether an individual studied computer science. The target variable here is directly generated from the computer science studying variable which is intuitive. An optimal predictor would take only the computer science course in order to predict the target. However when trying to achieve equalized odds, the predictor has to take the sex variable into account to adjust for the correlation between protected variable and independent variable.
\end{enumerate}

Using these scenarios, Hardt et al. \cite{hardt2016equality} show that it is possible to construct equal distributions for all variables such that the observable fairness indicators such as true and false positive rate are identical. Therefore if arguing that the perfect predictor is fair, one would allow the use of a variable directly indicating a protected variable or on the other hand, if arguing that the equalized odds predictor is fair, one would have to accept that it uses the protected variable to adjust for discrimination. This can be especially problematic if the use of such protected variables is forbidden, as the observational approach cannot decern wether the protected variable was used. Therefore Kilbertus et al. \cite{kilbertus2017avoiding} propose a new approach to detect and avoid discrimination in machine learning based on causal reasoning by Pearl \cite{pearl2009causality}.

\section{Literature: Discrimination detection and prevention with causal reasoning}

With his book on causality \cite{pearl2009causality}, Pearl lays the foundation for several causal inference approaches in the context of fairness. His approach of generating a set of confounding variables which explain causality between two variables $X$ and $Y$ allows the identification of causal graphs which in turn can be used to display causality as simple probability statements cannot \cite[p.~70]{pearl2009causality}. This approach aims at providing information on causation instead of correlation. Causation aims to infer how data was generated and how different variables were generated from other variables. Contrary to the observational approach, causal inference and causal reasoning require domain knowledge, i.e. one has to be able to reason about causality in the specific domain \cite{pearl2010intro}. The example in figure~\ref{fig:limobs} already uses a notation known as a causal graph which allows for a graphical visualization of the data generating process and is a directed acyclic graph. Contrary to the observational approach this now allows for assumptions about the relations and dependency structures within data and classifier. Kilbertus et al. \cite{kilbertus2017avoiding} differentiate between two new patterns of discrimination observable within such causal graphs namely unresolved discrimination and proxy discrimination. To find both structures, we always consider paths from protected variables to the classifier.

\subsubsection{Unresolved Discrimination}
For the first issue, consider all paths from the protected variable to the classifier to be discriminating or potential issues. The exceptions are paths which are blocked by a variable which is considered resolving. Blocking means that the variable lies on the path between first and last variable and can be the last variable itself. Resolving are variables which are partially generated by the protected variable but are not considered to be discriminating \cite{kilbertus2017avoiding}. Taking the example of figure~\ref{fig:limobs}, stack overflow is on the path from sex to both predictors but can be considered non-discriminating as the probability to use it is likely the same for software engineers regardless of genders. 

\subsubsection{Proxy Discrimination}
For the second issue, again consider all paths from the protected variable to the classifier, but assume all are not discriminating. The exception here is any path which contains a proxy, a variable which is generated by a protected variable and not considered a resolving variable . However the path then is only considered to exhibit potential proxy discrimination, which then has to be checked by doing an intervention \cite{kilbertus2017avoiding}. Taking the same example as for unresolved discrimination, candy crush is not a resolving variable as its sole purpose in the model is the approximation of sex using it. This approach, compared to detecting unresolved discrimination, is managed more easily as we only consider paths an issue if it is blocked by a proxy. Every proxy is labelled manually, therefore the number of problematic paths reduced.
\subsubsection{Preventing discrimination using causal inference}
After identifying discriminating variables and paths, one can alter the predictor with an optimization problem such that neither unresolved discrimination nor proxy discrimination are present. Starting out, one has to develop a structural equation model out of the dependency graph. Then all influences on the problematic variable have to be removed and all equations are combined into one equation defining the predictor $R$. Then, depending if proxies or resolving variables are considered, the proxy $P$ has to be intervened on or the protected variable $A$. The goal of the optimization problem is an equal probability distribution of $R$ for all possible interventions. An intervention is a process where all variables are fixated except for the problematic variable e.g. the proxy.

Nabi and Shpitser \cite{nabi2018fair} argue in a similar direction, as they gather disallowed paths and calculate a path specific effect which demonstrates the multiplier impact of a protected variable on a target variable. They then impose restrictions on their optimization problem such that for any problematic path the path specific effect is around 1, resulting in basically no effect.

\subsection{Counterfactual Fairness}
Kusner et al \cite{kusner2017counterfactual} propose a different approach which also has its roots in Pearls work \cite{pearl2009causality}. Again a structural equation model is developed which consitutes of assumptions made of the generation of underlying data. 
According to them a predictor is counterfactually fair if the distribution of predictor $\hat{Y}$ does not change if the protected variable $A$ is changed while all non causally dependent variables $X$ are held fixed. 
However they do not explicitly specify problematic paths or proxy variables but instead try to augment their data with additional latent causal variables $U$. These latent causal variables stem from a distribution of an underlying causal model. 
This model, similar to the causal inference approach by Kilbertus et al. \cite{kilbertus2017avoiding}, is again constructed by using domain knowledge yet they argue that it should be constantly re-evaluated when new information is available or new observations can only be explained with additional latent variables.
In an extension of \cite{kusner2017counterfactual}, Russell et al. \cite{russell2017worlds} provide an algorithm that no longer requires an exact causal model but instead allows for an approximation of counterfactual fairness across multiple causal models. These causal models could be different variants or originate from different statistical models and allow statisticians with less domain knowledge to make use of an approximate counterfactual fairness.
\newline
\newline 
The causal approach can identify discrimination in different ways and can develop models that avoid proxies of protected variables or include resolving variables. Yet the approach is complex and requires assumptions based on domain knowledge. Compared to the observational approach it does not just treat models as black boxes but there is another discipline within machine learning that attempts to make machine learning models interpretable and understandable which in turn can help to discover and avoid discrimination.

\section{Interpretability}

\textit{Interpretability} of machine learning models has yet to be properly defined \cite{lipton2016mythos}. Doshi-Velez and Kim provide their definition with \textit{the ability to explain or to present in understandable terms to a human}\cite{doshi2017towards}. Lipton \cite{lipton2016mythos} in an attempt to develop a definition therefore defines different contexts to which interpretability applies, specifically trust, causality, transferability, informativeness and fair and ethical decision-making. As the focus of our work lies on discrimination, we will mainly consider the fair and ethical decision making context. According to Goodman and Flaxman \cite{goodman2016european} this means that any model that should be interpretable needs to be able to be understood as well as articulated by humans. This of course relates to the \textit{right to explanation} in the GDPR as here it is also not defined what exactly the explanation entails. The legislation further supports the notion that some form of interpretability is needed particularly for very complex models. 

\subsection{Realizing interpretability}

Lipton \cite{lipton2016mythos} differentiates between different methods to make models interpretable which we will put into the fairness and discrimination context. He creates two not necessarily exclusive groups, transparency and post-hoc interpretability.

\subsubsection{Transparency:}
Transparency aims explain the workings of an algorithm by making its decisions or the entire learning process understandable.
The first option to provide transparency is to only use models which can be comprehended in their entirety by humans. This means that the model can be computed by a human in a small timeframe \cite{lipton2016mythos}. Ribeiro et al. \cite{ribeiro2016should} argue in a similar direction as they require a model to be presented to the user including images and text which further explain the model. However this presentation can get out of hand quickly if we consider any high dimensional data. Lou et al. \cite{lou2012intelligible} present another form of interpretability through transparency by arguing that a models features or decision points should be used as simple explanations. Specifically decision trees and regression models support this form of intelligibilty as e.g. in a tree a split could be all individuals from a bad neighborhood do not recieve a loan. Lipton \cite{lipton2016mythos} argues here that often weights and splits are largely based on which specific input variables where selected and the actual explanation in a real world context would be questionable at best. In the context of \textit{a right to explanation} and fairness, the transparency way of achieving interpretability largely requires very simple models with very limited data. As such models are often not capable of capturing complex relations, this would require sacrificing model accuracy for the benefit of being able to make the algorithm transparent to some extent.

\subsubsection{Post-hoc interpretability:}
\begin{figure}
    \centering
    \includegraphics[width=\textwidth]{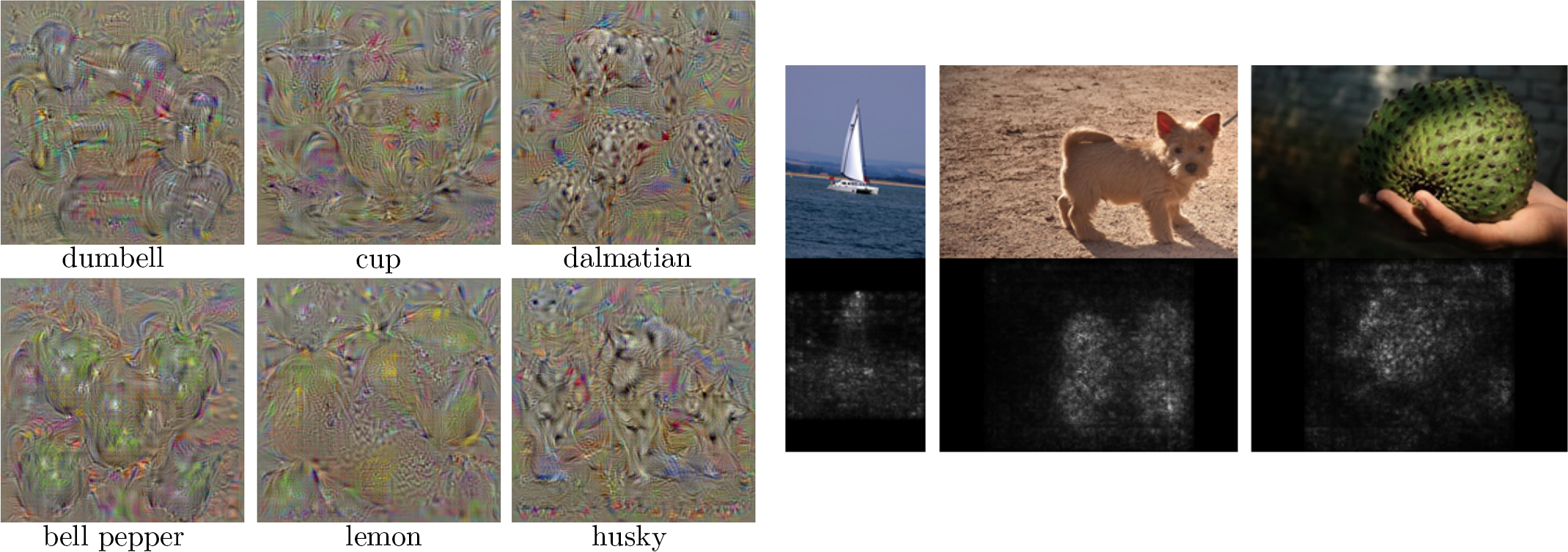}
    \caption{Class representations and Saliency maps \cite{simonyan2013deep}}
    \label{fig:viz}
\end{figure}

Post-hoc interpretability make use of already trained models and therefore do not explicitly explain the exact workings of an algorithm. Instead the goal here is to give useful information about the model in order to understand its decisions \cite{lipton2016mythos}.
One popular approach is, given a specific instance in a dataset which is to be discussed, to refer to similar instances i.e. instances which have similar activations in a neural network \cite{caruana1999case}. Similar to methods like case studies, this approach allows explanation which is very common for doctors when prescribing medicine (give sb. medicine because a very similar patient also got medicine and healed). This approach can be done after training as all model parameters are available to construct a table for similarity measurement. An alternative approach is the visualization or textualization of model parameters, which could also be part of the transparency approach of Ribeiro et al. \cite{ribeiro2016should}. 
Krening et al. \cite{krening2017learning} demonstrate textualization of decisions. They train a reinforcement learner which attempts to play super mario bros. by using and correctly classifying advice and warnings from human users. This sort of model could then also be used to output textual decisions when presented with a situation. McAuley and Leskovec \cite{mcauley2013hidden} train a latent-factor recommender system with review topics such that when returning a score on how likely the user will like a certain product, it can also give topical reasons for the user preference. Adding text to numbers indicating decisions can accordingly help to intuitively understand the decision making process of an algorithm.
Visualization to support interpretability is particularly important for algorithms analyizing images as a numeric representation would be less easily comprehensible. An early example of visualizing a convolutional neural networks features is presented by Simonyan et al. \cite{simonyan2013deep}. They generate images representative of a specific class to illustrate what the network \textit{thinks} such an object looks like. Furthermore they also generate saliency maps, indicating which areas of an input image were particularly impactful when deciding for a specific class. Both visualizations can help understand why the network provided such probability scores. An example can be found in Figure~\ref{fig:viz}.
The post-hoc form of interpretability can be particularly helpful for companies which already have an active machine learning classifier and now with the GDPR need to explain how decisions were made. 
\newline
\newline
Both basic forms of interpretability can help make avoid discriminating decisions or help to prove that an algorithm is not discriminating based on a protected attribute. Yet interpretability, apart from its lack of proper definition, has some issues. Firstly it, to some extent, requires machine learning models to never surpass human ability as it then automatically becomes more complex than is comprehensible. Secondly, because particularly post-hoc interpretability is also based on algorithms which are not transparent, the evaluations can be constructed such that they show no discriminating behavior yet the model still discriminates against a protected class. Here the person trying to understand the decision would again need to know how exactly the post-hoc interpretation was created \cite{lipton2016mythos}. With the \textit{right to explanation}, interpretability is a new necessity for any industries using machine learning to make decisions impacting individuals. However works like Hirsch et al. \cite{hirsch2017designing} already go one step further saying that algorithms not only have to be interpretable but contestable by affected individuals. This contestability includes possible modifications if the individual is rightly contesting the decision.

\section{Discussion}
Discrimination in machine learning has been an issue since companies started using algorithms to make important decisions. Already in 2006, an insurance company was sued for discriminating insurance scoring using an algorithm \cite{olddiscriminationexample}. New awareness has been achieved by the european union as new legislation now enforces a \textit{right to explanation} \cite{gdpr}.
Machine learning algorithms are made to discriminate, however they do not have to do so in a legal and ethical sense. The main reason for discrimination in machine learning seems to be the dataset an algorithm is trained on, if there is bias towards one group in the dataset, the algorithm will reflect it. If the dataset is unbalanced, the algorithm will reflect that as well. Using this information, discrimination can already be prevented by gathering unbiased and fair datasets which can also be supported by methods like causal inference. However considering that not every individual has only good intentions, the same can be done to achieve negative effect on other protected groups. For example, a potentially racist policeman could only ever check individuals of another ethnic group such that arrests only happen within that group. A dataset about arrests in an area would then reflect that unbalance. However often this kind of discrimination does not happen on purpose. Instead data scientists are simply not aware of the issues with their data and algorithm.

\subsection{Awareness}

Accordingly, one of the main issues regarding discrimination is the lack of awareness among practitioners. Simply by being aware, one can read into literature concerning the topic and can take steps to prevent discrimination in the own use case. Particularly when the algorithm one is writing strongly affects the life of an individual in cases such as arrest likelihood, insurance and credit lending. Awareness can be supported by a diverse workplace. With members of different protected groups present in a team, each individual is more sensitive of issues of another protected group and members of a discriminated group can immediately raise warnings. In the same way it may be necessary to have computer sciences students learn about discrimination, as it is often not part of their course. Especially knowing about the own impact on discriminating decisions and how to prevent them needs to be in the back of the mind of any person writing algorithms, which could strongly impact others lifes. Currently many students are oblivious to such issues and do not take actions to change this. Nonetheless being aware of issues is often not enough. Instead, domain knowledge is required. Machine learning algorithms and their implementations are constructed in a way that a practitioner can apply such algorithms to any dataset containing any number of variables of any type. Thus it is very simple to achieve a high accuracy. Yet there is no knowledge about the effect of different variables on each other, the meaning of variables and the potentially discriminating effect of variables. Specifically strategies like causal inference \cite{kilbertus2017avoiding} can only be used when one can reason about causality and build causal models, which requires both domain knowledge into the topic of inference and about the inner workings of the employed algorithm.

\subsection{Regarding prevention strategies}

Then again, even if one is aware of such issues, the proposed prevention strategies are lacking in many cases. The observational approaches such as equalized odds and equal opportunity are good starting points to evaluate discriminating behavior and to implement ethical ideals. Because theses approaches require information on protected variables, they are often not feasible in practice. Furthermore these approaches are very dependent on the used dataset and can be manipulated easily. Causal approaches on the other hand are very complex and require expertise in the domain the algorithm is applied to. The issues here is to build the right causal models and make the right assumptions about the generative process of the underlying data which is increasingly difficult, the more abstract the available data. Even multi-world approaches like Russell et al. \cite{russell2017worlds} cannot guarantee fairness and non-discrimination. Finally, all causal approaches require human decisions on which variables to classify as non-discriminating. Interpretability has to be another focus for practitioners as it is a way to create trust between affected individuals and algorithmic decisions. With the GDPR, interpretability is a requirement for any company making use of machine learning techniques, but this creates a conflict between the ability to solve very complex problems and the possibility to completely understand the decision-making process. Interpretability in any case supports the human review process when deciding whether an algorithm is discriminating against a protected group or not. Considering that all approaches show downsides, the best approach is likely to attempt to combine all three methods, making use of observable and causal procedures and facilitating them by highly interpretable algorithms.

\subsection{Future outlook}

The scientific community around machine learning is starting to put more effort into topics like fairness, discrimination and interpretability with total number of search results on the topic of fairness in machine learning having doubled since 2010 and interpretability having received almost 17500 new results \cite{scholar}. This of course is further sped up by new legislations like the GDPR in europe and new insitutions set up by the European Union such as the \textit{European Data Protection Supervisor}. These institutions will further improve legislations and likely properly define what an explanation for an algorithmic decision entails. It might be necessary to setup an institution that analyzes algorithms of companies and institutions and searches for discriminating behavior similar to what already exists for Vehicles and safety concerns or restaurants and hygene. Another issue for the future will be more and more complex algorithms where exact explanations are impossible. Hence explanations need to be simplified but according to works like Miller \cite{miller2017explanation} simple less accurate human-like explanations might be a better option anyway. Discrimination and fairness will depend on human intervention at least until ethics can be part of algorithms making humans, especially data scientists and machine learning experts, responsible to take action against any form of discrimination they can discover or that is reported. 

\section{Conclusion}
In this work we demonstrated discrimination as a problem in machine learning. We evaluated definitions and legal policies both on discrimination and algorithmic discrimination. This work then gave an overview on strategies to discover discriminating behavior and introductions to observational and causal approaches to prevent discrimination. We also demonstrated that interpretability can support processes to identify and prevent discrimination. Discrimination in machine learning has to be moved towards the center of attention especially in a diverse and global world such that decision-makers put resources into creating discrimination free algorithms. More work is needed in the field of fairness and interpretability and universities should start to require students in computer science and similar courses to take classes in ethics in computer science. The first step towards fairness is awareness. 

%
%
\bibliographystyle{splncs04}
\bibliography{bib}
\end{document}